\documentstyle[12pt,amssymb,epsfig,epsf]{article}

\renewcommand{\theequation}{\arabic{section}.\arabic{equation}}

\newcommand{\be}{\begin{equation}}
\newcommand{\ee}{\end{equation}}
\def \Tr {\mbox{Tr\,}}

\def \d  {\mbox{d}}

\begin{document}

\begin{titlepage}
\begin{flushright}
\begin{tabular}{l}
\end{tabular}
\end{flushright}

\vskip2cm
\begin{center}
  {\large \bf
     Baxter's Q-operator for the homogeneous XXX spin chain
  \\}

\vspace{2cm}
{\sc S.\'{E}. Derkachov}
\\[0.8cm]

\vspace*{0.1cm} {\it Department of Mathematics,
St.-Petersburg Technology Institute,\\ St.-Petersburg, Russia
                       } \\[0.2cm]

{\em Submitted to J.Phys.A:Math.Gen.}\\[1cm]

\vskip1.8cm
{\bf Abstract:\\[10pt]} \parbox[t]{\textwidth}{
Applying the Pasquier-Gaudin procedure we construct the
Baxter's Q-operator for the homogeneous XXX model as integral
operator in standard representation of $SL(2)$.
The connection between Q-operator and local Hamiltonians is discussed.
It is shown that operator of Lipatov's duality symmetry arises
naturally as leading term of the asymptotic expansion of
Q-operator for large values of spectral parameter.}
\vskip1cm

\end{center}

\end{titlepage}

\tableofcontents

\renewcommand{\theequation}{\thesubsection.\arabic{equation}}
\setcounter{equation}{0}

\section{Introduction.}

The modern approach to the theory of integrable systems is given by
the quantum inverse scattering method(QISM)~\cite{BE,F}.
In the framework of QISM, eigenstates
$|\lambda_1,...,\lambda_l\rangle$ are obtained
by the algebraic Bethe ansatz~(ABA) method as an excitations over
the vacuum state and the spectral problem is reduced to the set
of algebraic Bethe equations~(BE) for the parameters $\lambda_j$.
In fact the ABA is equivalent to the construction of the eigenfunctions
in a special representation as polynomials of some suitable variables.

The alternative approach is the method of Q-operator~\cite{B}
proposed by Baxter:there exists the operator $\hat{Q}(\lambda)$
which obeys the Baxter equation.
The set of the Bethe equations is
equivalent to the Baxter equation
for the eigenvalue $Q(\lambda)$ of the Q-operator.
This second order finite-difference equation is the simple
consequence of the Baxter relation for the
transfer matrix and the Q-operator~\cite{B}.

The ABA and method of Q-operator are equivalent when
eigenfunctions and therefore $Q(\lambda)$ are polynomials.
In more general "nonpolynomial" situation one could use the
method of Q-operator.
The Q-operator for the periodic Toda chain was
constructed in the work of Pasquier and Gaudin~\cite{G}.
The application of the Q-operator for the construction of
eigenstates with arbitrary complex values of conformal wights
in the case XXX spin chain was considered in the work of
Korchemsky and Faddeev~\cite{F}.
In the present paper we construct Q-operator for the homogeneous
XXX spin chain using the Pasquier-Gaudin procedure.

The presentation is organized as follows.
Section 2 introduces definitions and the standard facts
about Baxter equation and construction of local Hamiltonians.
In Section 3 we construct the Q-operator and study some properties
of the obtained Q-operator in the simplest case of homogeneous chain.
In Section 4 we obtain the connection between Q-operator
and local Hamiltonian.
In Section 5 we consider the asymptotic expansion of
Q-operator for large spectral parameter.
The operator of duality symmetry introduced by L.N.Lipatov~\cite{L} appears
naturally as leading term in this asymptotic.
Finally, in Section 6 we summarize.

\section{XXX spin chain}
\setcounter{equation}{0}

In this section we collect some basic facts about XXX spin chain.

\subsection{$R$-matrix and Yang-Baxter equation}
\setcounter{equation}{0}

The main object is the so called $R$ - matrix
which is the solution of the Yang-Baxter equation:
\begin{equation}
\label{YB}
R_{12}(\lambda) R_{13}(\lambda+\mu) R_{23}(\mu)=
R_{23}(\mu) R_{13}(\lambda+\mu) R_{12}(\lambda).
\end{equation}
The operator $R_{ij}(\lambda)$ depends on some complex
variable -- spectral parameter $\lambda$ and
two sets of $SL(2)$-generators $\vec{S}_i$ and
$\vec{S}_j$ acting in different vector spaces $V_i$ and $V_j$.

Fixing the representations of the spins $s_i$ and $s_j$ in the
vector spaces $V_i$ and $V_j$ we obtain the following $R$-matrices.
\begin{itemize}
\item
$s=1/2$ in space $V_i$ and arbitrary representation
$s$ for the $V_j$:
$$
R_{j}(\lambda) = \lambda + \frac{\eta}{2} +
\eta\cdot \vec{S}_j \vec\sigma.
$$
This $R$-matrix is used for the construction of the Lax $L$-operator:
\begin{equation}
\label{L}
L_i(\lambda)\equiv R(\lambda-\frac{\eta}{2}) =
\lambda +\eta\cdot \vec{S}_i \vec\sigma=
\left (\begin{array}{cc}
\lambda+\eta S_i & \eta S^{-}_i \\
 \eta\rm S^{+}_i & \lambda -\eta S_i \end{array} \right )
\end{equation}
\item
The equivalent representations $s$ in spaces
$V_i$ and $V_j$~\cite{KRS}:
\begin{equation}
\label{Fund}
R_{ij}(\lambda) = P_{ij}\cdot
\frac{\Gamma( J_{ij}+\eta \lambda)}
{\Gamma(J_{ij}-\eta \lambda)}
\ ;\ J_{ij}\cdot(J_{ij} -1) = L_{ij}
\end{equation}
where $P_{ij}$ is the permutation and
$L_{ij}$ is the "two-particle" Casimir in $V_i\otimes V_j$.
This fundamental $R$-matrix is the building block
for the construction of the local Hamiltonians.
\end{itemize}

\subsection{Baxter equation for XXX-model.}
\setcounter{equation}{0}

The "usual" quantum monodromy matrix $T(\lambda)$ is defined as
product of the $L$-matrices in the common
two-dimensional auxiliary space.
$T(\lambda)$ is a $2\times 2$ matrix with operator entries acting
in the quantum space $\otimes_{i=1}^{n} V_i$.
\begin{equation}
\label{monodr}
T (\lambda)\equiv
L_{1}(\lambda+ c_1) L_{2}(\lambda+c_2)...
L_{n}(\lambda+c_n) =
\left (\begin{array}{cc}
A(\lambda) & B(\lambda) \\
C(\lambda) & D(\lambda) \end{array} \right )
\end{equation}
The quantum transfer matrix $t(\lambda)$ is obtained
by taking trace of $T(\lambda)$ in the auxiliary space:
\begin{equation}
\label{trans}
t(\lambda)\equiv \Tr T(\lambda) =
A(\lambda) + D(\lambda)
\end{equation}
Due to the Yang-Baxter equation the family of operators
$t(\lambda)$ is commuting, its $\lambda$ expansion
begins with power $\lambda^n$ and provides $n-1$
commuting operators $Q_k$:
\begin{equation}
\label{Qk}
t(\lambda)\cdot t(\mu) =
t(\mu)\cdot t(\lambda)
\ ;\  t(\lambda) = 2\lambda^n+\sum_{k=0}^{n-2} Q_k \lambda^k.
\end{equation}
It is possible to show that transfer matrix $t(\lambda)$ is
$SL(2)$-invariant
$$
[\vec{S} , t(\lambda)] = 0
\ ;\  \vec{S} \equiv \sum_{k=1}^{n} \vec{S}_k.
$$
Therefore there exists the "full" set of $n$ commuting operators:
$n-1$ operators $Q_k$ and operator $S$.
Due to $SL(2)$-invariance the subspace of the eigenvectors
of operator $t(\lambda)$ with eigenvalue $\tau(\lambda)$ is
the $SL(2)$ - module generated
by highest weight vector $\Psi$,i.e. vector space spanned by
linear combinations of monomials in the $S^{+}$ applied
to vector $\Psi$.
The highest weight vector $\Psi$ is defined by
the equation $S^{-}\Psi = 0$.

We shall work in standard discrete-series representation of the group $SL(2)$:
$$
S \Psi(x)\equiv (cx+d)^{-2s} \Psi\biggl(\frac{ax+b}{cx+d}\biggr)
\ ;\  {\rm S}^{-1}=
\left (\begin{array}{cc}
a & b \\ c & d
\end{array} \right )
$$
where $SL(2)$-generators are realised as differential operators:
\be
\label{repr}
S_k = x_k\partial_k + s_k
\ ;\ S^{-}_k = -\partial_k
\ ;\ S^{+}_k = x^2_k\partial_k + 2 s_k x_k
\ee
acting in the space of polynomials of the variable $x_k$.
Here "spin" $s_k$ is arbitrary number.
In this representation the commuting operators $Q_k$ are
"local" differential operators acting in the space of
polynomials of the n variables $x_1,...,x_n$ and there
exists the vacuum vector $|0\rangle$:
$$
B(\lambda)|0\rangle = 0
\ ;\  A(\lambda)|0\rangle= \Delta_{+}(\lambda)|0\rangle
\ ;\  D(\lambda)|0\rangle= \Delta_{-}(\lambda)|0\rangle
$$
so that we can use the Algebraic Bethe Ansatz(ABA) method
and reduce the problem of the common diagonalization of the
operators $Q_k$ and $S$:
$$
t(\lambda)\Psi_l = \tau(\lambda) \Psi_l
\ ;\ S\Psi_l = (l+\sum_{k=1}^{n} s_k) \Psi_l
$$
to the solution of the Bethe equation~\cite{BE,F1}.
The vacuum vector $|0\rangle$ is the common highest vector
of the local representations of $SL(2)$:
$$
|0\rangle\equiv \prod_{k=1}^{n} |0\rangle_k
\ ;\  S^{-}_k|0\rangle_k=0
\ ;\  S_k|0\rangle_k= s_k|0\rangle_k,
$$
and
$$
L_{k}(\lambda+ c_k)|0\rangle_k=
\left (\begin{array}{cc}
\lambda + c_k +\eta s_k & 0 \\
 \cdots & \lambda + c_k - \eta s_k
\end{array} \right )|0\rangle_k
$$
so that
\be
\label{Delta}
\Delta_{\pm}(\lambda) \equiv
\prod_{k=1}^{n}(\lambda + c_k \pm\eta s_k).
\ee
Let us look now the eigenvector $\Psi_l$ in the form:
$$
\Psi_l\equiv|\lambda_1,...,\lambda_l\rangle
\equiv\prod_{j=1}^{l}{\rm C}(\lambda_j)|0\rangle
\ ;\  S|\lambda_1,...,\lambda_l\rangle =
(l+\sum_{k=1}^{n} s_k)|\lambda_1,...,\lambda_l\rangle.
$$
It is possible to show that vector $|\lambda_1,...,\lambda_l\rangle$ is
eigenvector of operator $t(\lambda)$ with eigenvalue:
\begin{equation}
\label{1}
\tau(\lambda)= \Delta_{+}(\lambda)
\prod_{j=1}^{l}\frac{(\lambda -\lambda_j+\eta)}{(\lambda-\lambda_j)}
+
\Delta_{-}(\lambda)
\prod_{j=1}^{l}\frac{(\lambda -\lambda_j-\eta)}{(\lambda-\lambda_j)}
\end{equation}
on condition that the parameters $\lambda_i$ obey the Bethe equations:
\begin{equation}
\label{2}
\prod_{j=1}^{l} (\lambda_i-\lambda_j+\eta)
\Delta_{+}(\lambda_i) =
\prod_{j=1}^{l} (\lambda_i-\lambda_j-\eta)
\Delta_{-}(\lambda_i).
\end{equation}
It appears also that Bethe vectors
$|\lambda_1,...,\lambda_l\rangle$ are the highest
weight vectors:
$$
S^{-}|\lambda_1,...,\lambda_l\rangle = 0.
$$
In the representation~(\ref{repr}) the highest weight
vector $\Psi_l$ is represented by
homogeneous, translation invariant polynomial
degree l($l= 0,1,2...$) of n variables $x_1,...,x_n$:
\begin{equation}
\label{S-}
\sum_{k=1}^{n} x_k \partial_k\Psi_l(x_1...x_n) =
l \Psi_l(x_1...x_n)
\ ;\ \sum_{k=1}^{n} \partial_k\Psi_l(x_1...x_n) = 0.
\end{equation}
One can obtain Bethe equation from the formula for $\tau(\lambda)$
by taking residue at $\lambda=\lambda_i$ and using the
fact that polynomial $\tau(\lambda)$ is regular at this point.
Finally we see that equations~(\ref{1},\ref{2}) are equivalent
to the Baxter equation for the polynomial $Q(\lambda)$:
\begin{equation}
\label{Baxter}
\tau(\lambda)Q(\lambda)=\Delta_{+}(\lambda)Q(\lambda+\eta) +
\Delta_{-}(\lambda)Q(\lambda-\eta)
\end{equation}
where
\be
\label{Q}
Q(\lambda)\equiv const\cdot\prod_{j=1}^{l}(\lambda-\lambda_j).
\ee

\subsection{Local Hamiltonians}
\setcounter{equation}{0}

Let us consider the homogeneous XXX-chain of equal spins:
$c_k=0$ and $s_k=s$ and fix the same representation $s$
in auxiliary space.
In this case the quantum monodromy matrix
$T_s(\lambda)$ is the product of the fundamental
$R$-matrices~(\ref{Fund}):
$$
T_s(\lambda)\equiv
R_{1}(\lambda) R_{2}(\lambda)...R_{n}(\lambda)
$$
The transfer matrix $t_s(\lambda)$ is obtained
by taking trace of $T_s(\lambda)$ in the auxiliary space
and due to the Yang-Baxter equation the families of operators
$t_s(\lambda)$ and $t(\lambda)$ are commuting:
$$
t_s(\lambda)\equiv \Tr_s T_s(\lambda)
\ ;\ t_s(\lambda)t_s(\mu) = t_s(\mu)t_s(\lambda)
\ ;\ t(\lambda)t_s(\mu) = t_s(\mu)t(\lambda).
$$
The $\lambda$ expansion of the $\log t_s(\lambda)$
provides Hamiltonians $H_k$:
\begin{equation}
\label{int}
H_k \equiv
\left.
\frac{1}{\eta}\frac{\partial^k}{\partial \lambda^k}
\log t_s(\lambda)\right|_{\lambda=0}
\ ;\ [ H_k, H_l ]= 0 \ ;\ [ H_k, Q_l ]= 0
\end{equation}
where the $k$-th operator describes the interaction between
$k+1$ nearest neighbours on the chain.
Due to the evident equalities~(see~(\ref{Fund})):
$$
R_{ij}(0) = P_{ij}
\ ;\ R^{\prime}_{ij}(0) = 2\eta P_{ij}\cdot\psi(J_{ij})
$$
one obtains the following expression for the first
"two-particle" Hamiltonian $H_1$:
$$
H_1 = \sum_{k=1}^{n} \hat{H}_{k-1,k}
\ ;\ \hat{H}_{k-1,k} = \frac{1}{\eta} P_{k-1,k} R^{\prime}_{k-1,k}=
2\cdot\psi( J_{k-1,k}),
$$
where $\psi(x)$ is logarithmic derivative of $\Gamma(x)$.
It is convenient to work with the "shifted" Hamiltonian:
\begin{equation}
\label{H}
H = \sum_{k=1}^{n} H_{k-1,k}
\ ;\ H_{k-1,k} = 2\cdot\psi( J_{k-1,k}) - 2 \psi(2s),
\end{equation}
where the "shift" constant is defined by the requirenent:
$$
H_{k-1,k}|0\rangle = 0.
$$
Let us calculate the eigenvalues of the operator
$H_{k-1,k}$. Operator $H_{k-1,k}$ is $SL(2)$-invariant
$$
[S^{\pm}_{k-1}+S^{\pm}_{k},H_{k-1,k}]=0
\ ;\  [S_{k-1}+S_{k},H_{k-1,k}]=0
$$
and its highest weight eigenfunctions $\Psi_l$
have the simple form in the representation~(\ref{repr}):
$$
(x_{k-1}\partial_{k-1}+x_{k}\partial_{k})\Psi_l = l \Psi_l
\ ,\ (\partial_{k-1}+\partial_{k})\Psi_l = 0
\Rightarrow
\Psi_l(x_{k-1},x_{k}) = (x_{k-1}-x_{k})^l.
$$
The two-particle Casimir $L_{k-1,k}$ is the
second order differential operator:
$$
L_{k-1,k}=-(x_{k-1}-x_k)^{2-2s}
\partial_{k-1}\partial_k (x_{k-1}-x_k)^{2s}
$$
and its eigenvalues $L_l$ and eigenvalues $J_l$ of
operator $J_{k-1,k}$ can be easily calculated:
$$
L_l = (2s+l)(2s+l-1)\ ;\ J_l = 2s+l.
$$
Finally we obtain the eigenvalues $H_l$ of the operator $H_{k-1,k}$:
$$
H_l = 2 \psi(2s+l) - 2 \psi(2s).
$$
In the representation~(\ref{repr}) the operator
$H_{k-1,k}$ can be realized as some "two-particle"
integral operator acting on the variables
$x_{k-1}$ and $x_k$:
\be
\label{Ham}
{\rm H}_{k-1,k} \Psi(x_{k-1},x_{k})=
-\int_{0}^{1} {\rm d}\alpha
\frac{\bar{\alpha}^{2s-1}}{\alpha}
\biggl[
\Psi(\bar{\alpha}x_{k-1}+\alpha x_{k},x_k)+
\Psi(x_{k-1},\alpha x_{k-1}+\bar{\alpha}x_{k})-
2 \Psi(x_{k-1},x_{k})\biggr].
\ee
Note that these integral operators arise naturally in QCD~\cite{We}.
To prove the equality~(\ref{Ham}) it is sufficient to show that
eigenvalues of integral operator coincide with the eigenvalues $H_l$:
$$
-2\int_{0}^{1} {\rm d}\alpha
\frac{\bar{\alpha}^{2s-1}}{\alpha}
\bigl[\bar{\alpha}^l-1 \bigr]= 2 [\psi(2s+l)-\psi(2s)].
$$
The expression for the eigenvalues of the full Hamiltonian $H$
can be found by the ABA method~\cite{H}:
$$
H = \frac{1}{\eta}\sum_{j=1}^{l}\frac{\partial}{\partial \lambda_j}
\log \frac{\lambda_j+\eta s}{\lambda_j-\eta s} =
\frac{1}{\eta}\sum_{j=1}^{l}\biggl[\frac{1}{\eta s - \lambda_j}
+\frac{1}{\eta s + \lambda_j}\biggr].
$$
It is possible to rewrite this expression in terms of the
$Q(\lambda)$-function~(\ref{Q}) as follows~\cite{F}:
\begin{equation}
\label{HQ}
H = \frac{Q^{\prime}(\eta s)}{\eta Q(\eta s)} -
\frac{Q^{\prime}(-\eta s)}{\eta Q(-\eta s)}
\end{equation}
There exists an additional commuting with transfer matrix
operator -- shift operator $P$:
\be
\label{P}
P \Psi(z_1,z_2...z_n) = \Psi(z_n,z_1...z_{n-1})
\ ;\ P = t_s(0).
\ee
Eigenvalues of the shift operator $P$ can be
found by the ABA method~\cite{H} also:
\be
\label{Pl}
P_l = \prod_{j=1}^{l}\frac{\lambda_j-\eta s}{\lambda_j+\eta s} =
\frac{Q(\eta s)}{Q(-\eta s)}.
\ee
In the next sections we shall construct the Baxter's Q-operator
and show that Baxter equation~(\ref{Baxter}) and
equations~(\ref{HQ},\ref{Pl}) arise from the corresponding
relations for the Q-operator.

\section{Baxter's  Q-operator.}
\setcounter{equation}{0}

The Baxter's $Q$ - operator is the operator
$\hat{Q}(\lambda)$ with the properties~\cite{B}:
\begin{itemize}
\item
$
t(\lambda)\hat{Q}(\lambda)=\Delta_{+}(\lambda)\hat{Q}(\lambda+\eta) +
\Delta_{-}(\lambda)\hat{Q}(\lambda-\eta)
$
\item
$\hat{Q}(\mu)\hat{Q}(\lambda)=
\hat{Q}(\lambda)\hat{Q}(\mu)
$
\item
$
t(\mu)\hat{Q}(\lambda)=
\hat{Q}(\lambda) t(\mu).
$
\end{itemize}
Operators $\hat{Q}(\lambda)$ and $t(\lambda)$
have the common set of eigenfunctions:
\be
\label{Eigen}
\hat{Q}(\lambda)\Psi = Q(\lambda)\cdot \Psi
\ ;\  t(\lambda)\Psi = \tau(\lambda)\cdot \Psi
\ee
and eigenvalues of these operators obey the
Baxter equation~(\ref{Baxter}).
Note that $Q$-function~(\ref{Q}) has the natural interpretation
as the eigenvalue of the Q-operator.

We construct the operator $\hat{Q}(\lambda)$ in
the standard representation of the group $SL(2)$
in the following form:
\begin{equation}
\label{q}
\hat{Q} (\lambda) \Psi(x) \equiv
\langle R Q(\lambda;x,z)|\Psi(z)\rangle
\ ;\ R Q(\lambda;z) \equiv z^{-2s}Q(\lambda;z^{-1})
\end{equation}
where $R$ is the transformation of inversion.
The scalar product here is the standard $SL(2)$-invariant
scalar product for functions of the one variable:
\be
\label{SL}
\langle \Psi(z)|\Phi(z)\rangle =
\int_{|z|\leq 1} D z \ \Psi(\bar{z})\Phi(z)
\ ;\ Dz \equiv \frac{2s-1}{\pi}
\frac{{\rm d}z{\rm d}\bar z}{(1-\bar{z} z)^{2-2s}}
\ee
and $z$ is "integration" or "dumb" variable.
In~(\ref{q}) the scalar product over all variables
$z_1...z_n$ is assumed.
The $SL(2)$-generators $S^{\pm}$ are conjugated
with respect to this scalar product:
$$
\langle \Psi|S^{\pm}\Phi\rangle =
\langle S^{\mp}\Psi|\Phi\rangle
\ ;\ \langle \Psi|S\Phi\rangle =
\langle S\Psi|\Phi\rangle.
$$
Using the evident identities:
$$
R \Phi(z) = z^{-2s} \Phi(z^{-1})
\ ;\ R S^{\pm}\Phi(z) = -S^{\mp} R \Phi(z)
\ ;\ R S\Phi(z) = - S R\Phi(z)
$$
we obtain the following rules for transposition:
\be
\label{conj}
\langle R Q(\lambda;z)|S^{\pm}\Psi(z)\rangle =
-\langle R S^{\pm}Q(\lambda;z)|\Psi(z)\rangle
\ ;\ \langle Q(\lambda;z)|S\Psi(z)\rangle =
-\langle R SQ(\lambda;z)|\Psi(z)\rangle
\ee
In fact the construction of the Q-operator repeates
the similar construction from the paper
Pasquier and Gaudin~\cite{G}.

The operator $t(\lambda)\equiv\Tr T(\lambda)$, where
$$
T(\lambda)\equiv
L_{1}(\lambda+ c_1)...L_{n}(\lambda+c_n),
$$
$$
L_{k}(\alpha_k)
=\eta\cdot\left (\begin{array}{cc}
\alpha_k+x_k\partial_k + s_k & -\partial_k \\
 x^2_k\partial_k + 2 s_k x_k & \alpha_k-x_k\partial_k-s_k
\end{array} \right )
\ ;\  \alpha_k=\frac{\lambda+ c_k}{\eta}
$$
is invariant with respect to transformation of
the local matrices $L_{k}$~\cite{B}:
$$
L_{k} \rightarrow
\bar{L}_{k}\equiv
N^{-1}_{k} L_{k} N_{k+1}
\ ;\  N_{n+1}\equiv  N_{1}
$$
where $N_{k}$ are the matrices with scalar elements.
Simple calculation shows that matrix elements of the
transformed matrix
$$
\bar{L}_{k}\equiv
N^{-1}_{k} L_{k} N_{k+1} =
\left (\begin{array}{cc}
\bar{L}_{k}^{11}&\bar{L}_{k}^{12}\\
\bar{L}_{k}^{12} & \bar{L}_{k}^{12}
\end{array} \right )
\ ;\ N_{k} =\left (\begin{array}{cc}
0 & 1 \\ -1 &  y_k
\end{array} \right )
$$
have the form
$$
\bar{L}_{k}^{11}=
-(x_k-y_k)^{1+\alpha_k-s_k}\partial_k (x_k-y_k)^{s_k-\alpha_k}
$$
$$
\bar{L}_{k}^{12}=
-(x_k-y_k)^{1+\alpha_k-s_k}(x_k-y_{k+1})^{1-\alpha_k-s_k}\partial_k
(x_k-y_k)^{s_k-\alpha_k}(x_k-y_{k+1})^{s_k+\alpha_k}
$$
$$
\bar{L}_{k}^{21}= \partial_k
\ ;\  \bar{L}_{k}^{22}=
(x_k-y_{k+1})^{1-\alpha_k-s_k}\partial_k
(x_k-y_{k+1})^{\alpha_k+s_k}.
$$
This expression for $\bar{L}$-operator suggests to
consider the function:
$$
\phi_k(\alpha_k;x_k;y_k,y_{k+1})\equiv
(x_k-y_k)^{\alpha_k-s_k}(x_k-y_{k+1})^{-\alpha_k-s_k}.
$$
The operators $\bar{L}_{k}^{ij}$ act on this
function as follows:
$$
\bar{L}_{k}^{11}\phi_k(\alpha_k)= (\alpha_k+s_k)\phi_k(\alpha_k+1)
\ ;\ \bar{L}_{k}^{12}\phi_k(\alpha_k)=0
$$
$$
\bar{L}_{k}^{22}\phi_k(\alpha_k)=
(\alpha_k-s_k)\phi(\alpha_k-1).
$$
Let us fix the dependence on the $x$-variables in the
kernel of operator $\hat{Q}(\lambda)$ in the form:
$$
Q(\lambda;x) \leftrightarrow \prod_{k=1}^{n}
\phi(\alpha_k;x_k;y_k,y_{k+1})
\ ;\ \eta \alpha_k = \lambda+c_k,
$$
where  $\{y_i\}$ - is the set of arbitrary parameters now.
Then we have:
$$
t(\lambda) Q(\lambda;x;y)=
\eta^n\cdot \Tr
\prod_{k=1}^{n}\bar{L}_{k}\phi_k(\alpha_k)=
\prod_{k=1}^{n}
\left (\begin{array}{cc}
(\alpha_k+s_k)\phi_k(\alpha_k+1) & 0 \\ \ldots  &
(\alpha_k-s_k)\phi_k(\alpha_k-1)
\end{array} \right ).
$$
After multiplication of these triangular matrices and
calculation of the trace we obtain the "right"
Baxter's relation:
$$
t(\lambda) Q(\lambda;x)=
\Delta_{+}(\lambda) Q(\lambda+\eta;x)
+ \Delta_{-}(\lambda) Q(\lambda-\eta;x).
$$
Next step we fix the dependence on
the $z$-variables in the kernel of operator
$\hat{Q}(\lambda)$ to obtain the "left" Baxter's relation:
$$
Q(\lambda;x,z) t(\lambda)=
\Delta_{+}(\lambda) Q(\lambda+\eta;x,z)
+ \Delta_{-}(\lambda) Q(\lambda-\eta;x,z).
$$
The rules~(\ref{conj}) allow to move $SL(2)$-generators
from the function $\Psi(z)$ to the kernel of the Q-operator:
$$
\langle R Q (\lambda;x,z)|L_1...L_n \Psi(z)\rangle =
\langle R L^{\prime}_n...L^{\prime}_1
Q (\lambda;x,z)| \Psi(z)\rangle
$$
where
$$
L_{k}^{\prime}\equiv \eta\cdot\left (\begin{array}{cc}
\alpha_k- S_k & -S^{-}_k \\
 -S^{+}_k & \alpha_k+S_k \end{array} \right )
=\bigl[\sigma_2\cdot L_{k} \sigma_2\bigr]^{t},
$$
and $t$ means transposition.
The trace of the product of $L^{\prime}$-matrices can be
calculated
$$
\Tr L^{\prime}_n...L^{\prime}_1 =
\Tr \bigl[L_1...L_n\bigr]^{t} =
\Tr L_1...L_n,
$$
and finally we obtain:
$$
\langle R Q (\lambda;x,z)|\Tr \bigl[L_1...L_n\bigr] \Psi(z)\rangle =
\langle R  \Tr\bigl[L_1...L_n\bigr]
Q (\lambda;x,z)| \Psi(z)\rangle
$$
Therefore the dependence of the kernel $Q(\lambda;x,z)$
on the $x$- and $z$-variables have the same form.

In the sequel we shall concentrate on the case of
homogeneous XXX-chain.

\subsection{Q-operator for the homogeneous XXX-chain}
\setcounter{equation}{0}

In this section we consider the homogeneous XXX-chain of equal spins:
$\ c_i = 0\ ,\ s_i=s$.
The kernel:
$$
Q(\lambda;x,z)\equiv
(-1)^{-2sn}\prod_{k=1}^{n} (x_k-z_k)^{-\frac{\eta s-\lambda}{\eta}}
(x_k-z_{k+1})^{-\frac{\eta s+\lambda}{\eta}}
$$
has the "true" $x$- and $z$-dependences and therefore
$Q$-operator can be defined as follows:
$$
\hat{Q} (\lambda) \Psi(x) =
\langle R Q(\lambda;x,z)|\Psi(z)\rangle=
\prod_{k=1}^{n} \langle (1-z_k x_{k})^{-\frac{\eta s-\lambda}{\eta}}
(1-z_{k}x_{k-1})^{-\frac{\eta s+\lambda}{\eta}}|\Psi(z)\rangle.
$$
There exists some useful integral representations for
obtained Q-operator.

\subsection{The $\alpha$-representation for the Q-operator}
\setcounter{equation}{0}

Let us consider the $Q$-operator:
$$
\hat{Q} (\lambda) \Psi(x_1...x_n) \equiv
\prod_{k=1}^{n} \langle
(1-x_{k-1}z_{k})^{-\frac{\eta s+\lambda}{\eta}}
(1-x_{k}z_k)^{-\frac{\eta s-\lambda}{\eta}}|\Psi(z_1...z_n)\rangle
$$
and transform the $z_k$-integral using the following identity:
$$
\int_{|z_k|\leq 1} Dz_k\ (1-x_k\bar{z}_k)^{-a}
(1-x_{k-1}\bar{z}_k)^{-b}\Psi(z_k)=
$$
\begin{equation}
\label{id}
=\frac{\Gamma(2s)}{\Gamma(a)\Gamma(b)}\cdot
\int_{0}^{1} \d\alpha
\alpha^{a-1}(1-\alpha)^{b-1}
\Psi[\alpha x_k+(1-\alpha)x_{k-1}\bigr] \ ;\ a+b=2s.
\end{equation}
To prove this identity we use the Feynman formula:
\be
\label{F}
\frac{1}{A^a B^b} = \frac{\Gamma(a+b)}{\Gamma(a)\Gamma(b)}\cdot
\int_{0}^{1} \d\alpha \alpha^{a-1}(1-\alpha)^{b-1}
\frac{1}{\bigl[\alpha A+(1-\alpha)B\bigr]^{a+b}}
\ee
and transform the product:
$$
(1-x_k\bar{z}_k)^{-a}
(1-x_{k-1}\bar{z}_k)^{-b}=
\frac{\Gamma(a+b)}{\Gamma(a)\Gamma(b)}\cdot
\int_{0}^{1} \d\alpha
\frac{\alpha^{a-1}(1-\alpha)^{b-1}}
{\bigl[1-(\alpha x_k+(1-\alpha)x_{k-1})\bar{z}_k\bigr]^{2s}}.
$$
The remaining $z$-integral can be easily calculated:
$$
\int_{|z_k|\leq 1} Dz_k\ (1-x\bar{z}_k)^{-2s}\Psi(z_k) = \Psi(x)
\ ;\ x \equiv \alpha x_k+(1-\alpha)x_{k-1}.
$$
Finally we obtain the useful integral representation
($\alpha$-representation) for the $Q$-operator:
\begin{equation}
\label{alpha}
\hat{Q}(\lambda) \Psi(x) \equiv
\prod_{k=1}^{n}
\Gamma(\lambda;s)\int_{0}^{1} {\rm d}\alpha_k
\alpha_k^{\frac{\eta s-\lambda}{\eta}-1}
\bar{\alpha}_k^{\frac{\eta s+\lambda}{\eta}-1}
\Psi\bigl[...\alpha_k x_k+\bar{\alpha}_k x_{k-1}...\bigr],
\end{equation}
where $\bar{\alpha} \equiv 1-\alpha$ and
$$
\Gamma(\lambda;s)\equiv\frac{\Gamma(2s)}{\Gamma(s+\lambda\eta^{-1})
\Gamma(s-\lambda\eta^{-1})}.
$$
Let us consider the eigenvalue problem for the Q-operator:
$$
\hat{Q}(\lambda)\Psi(x) = Q(\lambda) \Psi(x),
$$
where polynomial $\Psi(x)$ belongs to the space
of homogeneous polynomials degree $l$~(\ref{S-}):
\be
\label{l}
\Psi(x) = \sum_{p}\Psi_{p_1...p_n} x_1^{p_1}...x_n^{p_n}
\ ;\ p_1+p_2+...+p_n = l\ ;\ l=0,1,2...
\ee
The Q-operator transforms polynomial $\Psi(x)$ to
homogeneous polynomial degree $l$ whose coefficients are
polynomials in $\lambda$ degree $l$.
Therefore eigenvalues $Q(\lambda)$ of the Q-operator
are polynomials in $\lambda$ degree $l$.

For the proof we use obtained $\alpha$-representation.
Let us consider the action of Q-operator on
polynomial $\Psi(x)$:
$$
\hat{Q}(\lambda) \Psi(x) \equiv
\sum_{p}\Psi_{p_1...p_n}\prod_{k=1}^{n}
\Gamma(\lambda;s)\int_{0}^{1} {\rm d}\alpha_k
\alpha_k^{\frac{\eta s-\lambda}{\eta}-1}
\bar{\alpha}_k^{\frac{\eta s+\lambda}{\eta}-1}
\bigl[\alpha_k x_k+\bar{\alpha}_k x_{k-1}\bigr]^{p_k}.
$$
The expression for the $\alpha_k$-integral have the form:
$$
\Gamma(\lambda;s)\int_{0}^{1} {\rm d}\alpha_k
\alpha_k^{\frac{\eta s-\lambda}{\eta}-1}
\bar{\alpha}_k^{\frac{\eta s+\lambda}{\eta}-1}
\bigl[\alpha_k x_k+\bar{\alpha}_k x_{k-1}\bigr]^{p_k}=
\sum_{m=0}^{p_k} C_{p_k,m} x_k^m x_{k-1}^{p_k-m},
$$
where coefficients $C_{p_k,m}$
$$
C_{p_k,m}= \frac{p_k!}{m!(p_k-m)!}
\frac{\Gamma(2s)}{\Gamma(2s+p_k)}
\frac{\Gamma(s-\lambda\eta^{-1}+m)}{\Gamma(s-\lambda\eta^{-1})}
\frac{\Gamma(s+\lambda\eta^{-1}+p_k-m)}{\Gamma(s+\lambda\eta^{-1})}
$$
are polynomials in $\lambda$ degree $p_k$ because of evident equality:
$$
\frac{\Gamma(a+m)}{\Gamma(a)}= a(a+1)...(a+m-1).
$$
There are similar expressions for the remaining $\alpha$-integrals
and we obtain that Q-operator transforms polynomial $\Psi(x)$ to
homogeneous polynomial degree $l$ whose coefficients are
polynomials in $\lambda$ degree $p_1+p_2+...p_n=l$.

There exists some another useful representation for the
Q-operator(t-representation):
\be
\label{t}
\hat{Q} (\lambda) \Psi(x) \equiv
\prod_{k=1}^{n} \frac{\Gamma(\lambda;s)}
{(x_k-x_{k-1})^{2s-1}}\int_{x_{k-1}}^{x_{k}} {\rm d}t_k
(t_k-x_{k-1})^{\frac{\eta s-\lambda}{\eta}-1}
(x_k-t_k)^{\frac{\eta s+\lambda}{\eta}-1}\Psi[...t_k...].
\ee
This formula is obtained from the~(\ref{alpha}) by
the following change of variables:
$$
t_k = \alpha_k x_k+\bar{\alpha}_k x_{k-1}.
$$

\subsection{$SL(2)$-invariance of the Q-operator. Commutativity}
\setcounter{equation}{0}

We shall prove the two important properties of obtained Q-operator:
$SL(2)$-invariance and commutativity.
Let us begin from the $SL(2)$-invariance:
$$
S \hat{Q}(\lambda)\Psi(x) = \hat{Q}(\lambda) S\Psi(x)
\ ;\ S \Psi(x)\equiv (cx+d)^{-2s} \Psi(Sx)
\ ;\ S x\equiv \frac{ax+b}{cx+d}.
$$
The simplest way is to use the representation~(\ref{t}).
We start from $S\hat{Q}(\lambda)$:
$$
S \hat{Q}(\lambda) \Psi(x) \sim
(Sx_k-Sx_{k-1})^{-2s+1}\cdot
\int_{Sx_{k-1}}^{Sx_{k}} {\rm d}t
(t-Sx_{k-1})^{\frac{\eta s-\lambda}{\eta}-1}
(Sx_k-t)^{\frac{\eta s+\lambda}{\eta}-1}
\Psi(t)
$$
and make the change of variable in $t$-integral:
$$
t = S\tau = \frac{a\tau+b}{c\tau+d}
\ ;\ S\tau-Sx = \frac{\tau - x}{(c\tau+d)(cx+d)}
\ ;\ \d t = \frac{\d\tau}{(c\tau+d)^2}.
$$
After this change of variable the $t$-integral is transformed to the
$\tau$-integral of required form:
$$
(x_k-x_{k-1})^{-2s+1}\cdot
\int_{x_{k-1}}^{x_{k}} {\rm d}\tau
(\tau-x_{k-1})^{\frac{\eta s-\lambda}{\eta}-1}
(x_k-\tau)^{\frac{\eta s+\lambda}{\eta}-1}
(c \tau+d)^{-2s}\Psi(S\tau) \sim \hat{Q}(\lambda) S\Psi(x).
$$
It is worth to emphasize that all factors like $(c x_k+d)^{-2s}$ are
cancelled in the whole product.

The second important property of Q-operator is commutativity:
\be
\label{qq}
\hat{Q}(\mu)\hat{Q}(\lambda) =
\hat{Q}(\lambda)\hat{Q}(\mu).
\ee
It follows that there exists a
unitary operator $U$ independent on $\lambda$ which diagonalizes
$\hat{Q}(\lambda)$ simultaneously for all values of $\lambda$ and
therefore due to the Baxter relation operators
$\hat{Q}(\lambda)$ and $t(\mu)$ commute also:
\be
\label{qt}
t(\mu)\hat{Q}(\lambda)=
\hat{Q}(\lambda) t(\mu).
\ee
It is useful to visualize the Q-operator itself
and the product of two Q-operators as shown in
figure: the line with index $a$ between the points $x$ and $z$
represents the function $(1-xz)^{-a}$ where
$\ a = \frac{\eta s -\lambda}{\eta}$ and $b = \frac{\eta s -\mu}{\eta}$.
The integration~(\ref{SL}) in any four-point vertex is supposed.

\hspace{20 mm}

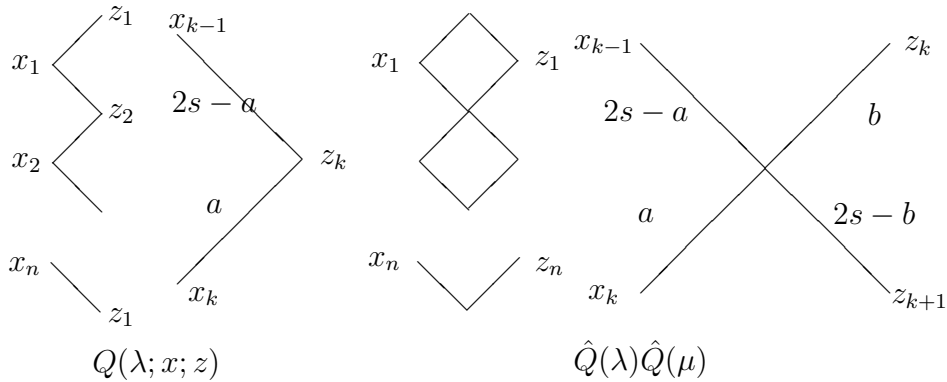
\unitlength 0.70mm
\linethickness{0.4pt}
\begin{picture}(176.66,75.67)
\put(35.67,71.66){\line(1,-1){23.67}}
\put(59.33,48.00){\line(-1,-1){23.67}}
\put(123.66,70.00){\line(1,-1){23.67}}
\put(147.33,46.33){\line(-1,-1){23.67}}
\put(171.00,70.00){\line(-1,-1){23.67}}
\put(147.33,46.33){\line(1,-1){23.67}}
\put(40.66,22.00){\makebox(0,0)[cc]{$x_{k}$}}
\put(39.67,73.67){\makebox(0,0)[cc]{$x_{k-1}$}}
\put(42.67,58.66){\makebox(0,0)[cc]{$2s-a$}}
\put(42.67,39.00){\makebox(0,0)[cc]{$a$}}
\put(65.33,48.00){\makebox(0,0)[cc]{$z_{k}$}}
\put(31.66,9.00){\makebox(0,0)[cc]{$Q(\lambda;x;z)$}}
\put(116.66,69.66){\makebox(0,0)[cc]{$x_{k-1}$}}
\put(116.66,22.33){\makebox(0,0)[cc]{$x_{k}$}}
\put(176.33,69.66){\makebox(0,0)[cc]{$z_k$}}
\put(176.66,21.33){\makebox(0,0)[cc]{$z_{k+1}$}}
\put(124.66,57.00){\makebox(0,0)[cc]{$2s-a$}}
\put(124.66,37.33){\makebox(0,0)[cc]{$a$}}
\put(168.00,56.66){\makebox(0,0)[cc]{$b$}}
\put(168.00,37.66){\makebox(0,0)[cc]{$2s-b$}}
\put(123.66,10.00){\makebox(0,0)[cc]{$\hat{Q}(\lambda)\hat{Q}(\mu)$}}
\put(91.00,75.67){\line(-1,-1){9.33}}
\put(81.67,66.33){\line(1,-1){9.33}}
\put(91.00,57.00){\line(-1,-1){9.33}}
\put(81.67,47.67){\line(1,-1){9.33}}
\put(81.33,28.67){\line(1,-1){9.33}}
\put(21.33,75.33){\line(-1,-1){9.33}}
\put(12.00,66.00){\line(1,-1){9.33}}
\put(21.33,56.67){\line(-1,-1){9.33}}
\put(12.00,47.33){\line(1,-1){9.33}}
\put(11.67,28.33){\line(1,-1){9.33}}
\put(7.00,66.00){\makebox(0,0)[cc]{$x_1$}}
\put(7.00,47.33){\makebox(0,0)[cc]{$x_2$}}
\put(6.67,27.67){\makebox(0,0)[cc]{$x_n$}}
\put(25.00,75.33){\makebox(0,0)[cc]{$z_1$}}
\put(25.00,56.67){\makebox(0,0)[cc]{$z_2$}}
\put(25.00,18.00){\makebox(0,0)[cc]{$z_1$}}
\put(91.33,75.67){\line(1,-1){9.00}}
\put(100.33,66.67){\line(-1,-1){9.33}}
\put(91.00,57.33){\line(1,-1){9.33}}
\put(75.33,66.33){\makebox(0,0)[cc]{$x_1$}}
\put(75.00,28.67){\makebox(0,0)[cc]{$x_n$}}
\put(106.00,66.67){\makebox(0,0)[cc]{$z_1$}}
\put(106.33,27.67){\makebox(0,0)[cc]{$z_n$}}
\put(91.00,38.67){\line(1,1){9.33}}
\put(90.67,19.33){\line(1,1){10.00}}
\end{picture}

\hspace{5 mm}

Let us consider the product
$\hat{Q}(\lambda)\hat{Q}(\mu)$ and
the corresponding kernel:
$$
\langle Q(\lambda;x;y)|Q(\mu;y;z)\rangle\equiv
\prod_{k=1}^{n}
\langle (1 - x_{k-1} y_k)^{\frac{-\lambda-\eta s}{\eta}}
(1 - x_k y_k)^{\frac{\lambda-\eta s}{\eta}}|
(1- y_k z_{k+1})^{\frac{\mu-\eta s}{\eta}}
(1-y_k z_k)^{\frac{-\mu-\eta s}{\eta}}\rangle.
$$
The "mechanism" of commutativity is shown in figure~\cite{G}:

\hspace{20 mm}

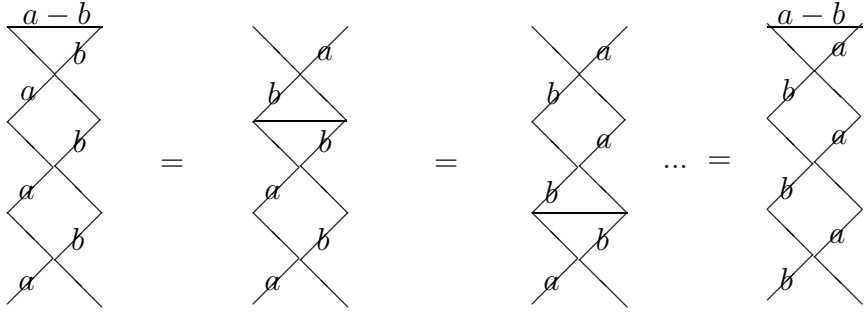
\unitlength 0.70mm
\linethickness{0.4pt}
\begin{picture}(177.67,68.00)
\put(24.33,56.67){\line(1,1){9.00}}
\put(33.33,65.67){\line(-1,0){18.00}}
\put(15.33,65.67){\line(1,-1){9.00}}
\put(24.33,56.67){\line(-1,-1){9.00}}
\put(15.33,47.67){\line(1,-1){8.67}}
\put(24.00,39.00){\line(-1,-1){8.67}}
\put(15.33,30.33){\line(1,-1){8.67}}
\put(24.00,21.67){\line(-1,-1){8.67}}
\put(24.33,56.67){\line(1,-1){8.67}}
\put(33.00,48.00){\line(-1,-1){8.67}}
\put(24.33,39.33){\line(1,-1){9.00}}
\put(33.33,30.33){\line(-1,-1){9.00}}
\put(24.33,21.33){\line(1,-1){9.00}}
\put(24.67,68.00){\makebox(0,0)[cc]{$a-b$}}
\put(19.33,53.00){\makebox(0,0)[cc]{$a$}}
\put(19.00,34.33){\makebox(0,0)[cc]{$a$}}
\put(19.00,17.00){\makebox(0,0)[cc]{$a$}}
\put(29.00,60.67){\makebox(0,0)[cc]{$b$}}
\put(29.00,44.00){\makebox(0,0)[cc]{$b$}}
\put(28.67,25.33){\makebox(0,0)[cc]{$b$}}
\put(71.00,56.67){\line(1,1){9.00}}
\put(62.00,65.67){\line(1,-1){9.00}}
\put(71.00,56.67){\line(-1,-1){9.00}}
\put(62.00,47.67){\line(1,-1){8.67}}
\put(70.67,39.00){\line(-1,-1){8.67}}
\put(62.00,30.33){\line(1,-1){8.67}}
\put(70.67,21.67){\line(-1,-1){8.67}}
\put(71.00,56.67){\line(1,-1){8.67}}
\put(79.67,48.00){\line(-1,-1){8.67}}
\put(71.00,39.33){\line(1,-1){9.00}}
\put(80.00,30.33){\line(-1,-1){9.00}}
\put(71.00,21.33){\line(1,-1){9.00}}
\put(66.00,53.00){\makebox(0,0)[cc]{$b$}}
\put(65.67,34.33){\makebox(0,0)[cc]{$a$}}
\put(65.67,17.00){\makebox(0,0)[cc]{$a$}}
\put(75.67,60.67){\makebox(0,0)[cc]{$a$}}
\put(75.67,44.00){\makebox(0,0)[cc]{$b$}}
\put(75.33,25.33){\makebox(0,0)[cc]{$b$}}
\put(124.00,56.67){\line(1,1){9.00}}
\put(133.00,30.33){\line(-1,0){18.00}}
\put(115.00,65.67){\line(1,-1){9.00}}
\put(124.00,56.67){\line(-1,-1){9.00}}
\put(115.00,47.67){\line(1,-1){8.67}}
\put(123.67,39.00){\line(-1,-1){8.67}}
\put(115.00,30.33){\line(1,-1){8.67}}
\put(123.67,21.67){\line(-1,-1){8.67}}
\put(124.00,56.67){\line(1,-1){8.67}}
\put(132.67,48.00){\line(-1,-1){8.67}}
\put(124.00,39.33){\line(1,-1){9.00}}
\put(133.00,30.33){\line(-1,-1){9.00}}
\put(124.00,21.33){\line(1,-1){9.00}}
\put(119.00,53.00){\makebox(0,0)[cc]{$b$}}
\put(118.67,34.33){\makebox(0,0)[cc]{$b$}}
\put(118.67,17.00){\makebox(0,0)[cc]{$a$}}
\put(128.67,60.67){\makebox(0,0)[cc]{$a$}}
\put(128.67,44.00){\makebox(0,0)[cc]{$a$}}
\put(128.33,25.33){\makebox(0,0)[cc]{$b$}}
\put(46.67,39.33){\makebox(0,0)[cc]{$\ =\ $}}
\put(98.67,39.33){\makebox(0,0)[cc]{$\ =\ $}}
\put(62.33,47.67){\line(1,0){17.33}}
\put(168.67,57.34){\line(1,1){9.00}}
\put(177.67,65.67){\line(-1,0){18.00}}
\put(159.67,66.34){\line(1,-1){9.00}}
\put(168.67,57.34){\line(-1,-1){9.00}}
\put(159.67,48.34){\line(1,-1){8.67}}
\put(168.34,39.67){\line(-1,-1){8.67}}
\put(159.67,31.00){\line(1,-1){8.67}}
\put(168.34,22.34){\line(-1,-1){8.67}}
\put(168.67,57.34){\line(1,-1){8.67}}
\put(177.34,48.67){\line(-1,-1){8.67}}
\put(168.67,40.00){\line(1,-1){9.00}}
\put(177.67,31.00){\line(-1,-1){9.00}}
\put(168.67,22.00){\line(1,-1){9.00}}
\put(168.00,68.00){\makebox(0,0)[cc]{$a-b$}}
\put(163.67,53.67){\makebox(0,0)[cc]{$b$}}
\put(163.34,35.00){\makebox(0,0)[cc]{$b$}}
\put(163.34,17.67){\makebox(0,0)[cc]{$b$}}
\put(173.34,61.34){\makebox(0,0)[cc]{$a$}}
\put(173.34,44.67){\makebox(0,0)[cc]{$a$}}
\put(173.00,26.00){\makebox(0,0)[cc]{$a$}}
\put(146.34,40.00){\makebox(0,0)[cc]{$\ ...\ =\ $}}
\end{picture}
\hspace{5 mm}

and is grounded on the "local" identity:
\begin{equation}
\label{loc}
\langle (1 - x_{k-1} y)^{-2s+a}(1- x_k y)^{-a}|
(1- y z_{k})^{-b}(1-y z_{k+1})^{-2s+b}\rangle
\cdot (1- z_k x_{k-1})^{a-b}=
\end{equation}
$$
=
\langle
(1- x_{k-1} y)^{-2s+b}(1- x_k y)^{-b}|
(1 - y z_{k})^{-a}(1-y z_{k+1})^{-2s+a}\rangle
\cdot (1-z_{k+1}x_{k})^{a-b}.
$$
The graphic representation of this identity is shown
in figure ($a,b$ are arbitrary parameters):

\hspace{20 mm}

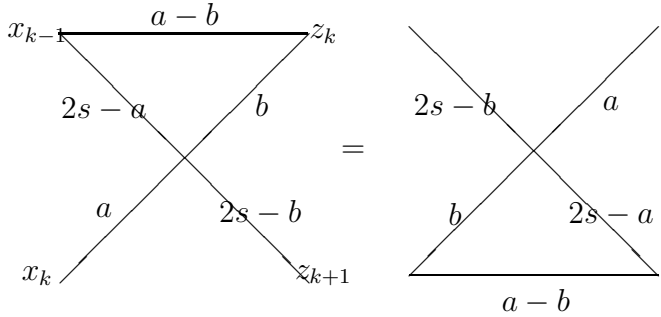
\unitlength 0.70mm
\linethickness{0.4pt}
\begin{picture}(126.67,71.67)
\put(79.33,69.67){\line(1,-1){23.67}}
\put(103.00,46.00){\line(-1,-1){23.67}}
\put(126.67,69.67){\line(-1,-1){23.67}}
\put(103.00,46.00){\line(1,-1){23.67}}
\put(13.00,68.34){\line(1,-1){23.67}}
\put(36.67,44.67){\line(-1,-1){23.67}}
\put(60.34,68.34){\line(-1,-1){23.67}}
\put(36.67,44.67){\line(1,-1){23.67}}
\put(68.67,46.00){\makebox(0,0)[cc]{$=$}}
\put(79.33,22.33){\line(1,0){47.33}}
\put(8.67,22.33){\makebox(0,0)[cc]{$x_{k}$}}
\put(63.00,22.33){\makebox(0,0)[cc]{$z_{k+1}$}}
\put(12.67,68.33){\line(1,0){47.33}}
\put(8.67,68.33){\makebox(0,0)[cc]{$x_{k-1}$}}
\put(63.00,68.33){\makebox(0,0)[cc]{$z_k$}}
\put(21.33,34.67){\makebox(0,0)[cc]{$a$}}
\put(21.33,54.00){\makebox(0,0)[cc]{$2s-a$}}
\put(51.33,54.67){\makebox(0,0)[cc]{$b$}}
\put(51.00,34.33){\makebox(0,0)[cc]{$2s-b$}}
\put(36.67,71.67){\makebox(0,0)[cc]{$a-b$}}
\put(103.33,17.33){\makebox(0,0)[cc]{$a-b$}}
\put(117.67,55.00){\makebox(0,0)[cc]{$a$}}
\put(88.00,54.67){\makebox(0,0)[cc]{$2s-b$}}
\put(88.00,34.00){\makebox(0,0)[cc]{$b$}}
\put(117.67,34.00){\makebox(0,0)[cc]{$2s-a$}}
\end{picture}
\hspace{5 mm}

The proof of the equality~(\ref{loc}) can be found in Appendix.

\subsection{Eigenvalues of the Q - operator for $n=2$}
\setcounter{equation}{0}

The case $n=2$ is the simplest one:
$$
t(\lambda) = 2\lambda^2 + 2 \eta^2 \vec{S}_1 \vec{S}_2 =
2\lambda^2 -2\eta^2 s(s-1) + \eta^2 L.
$$
There exists only one integral of motion -- two-particle
Casimir: $\ L \equiv(\vec{S}_1 +\vec{S}_2)^2$.
Its highest weight eigenfunctions have the form:
$$
\Psi_l(x_1,x_2)=(x_1-x_2)^l
\ ;\ L \Psi_l = (l+2s)(l+2s-1)\Psi_l.
$$
Due to $SL(2)$-invariance these functions are
eigenfunctions for the Q-operator also.
Let us calculate the eigenvalue $Q_l(\lambda)$:
$$
\hat{Q}(\lambda)\Psi_l = Q_l(\lambda)\cdot \Psi_l.
$$
The simplest way is to use the $\alpha$-representation:
$$
\hat{Q} (\lambda) \Psi_l \equiv
\Gamma^2(\lambda;s)
\int_{0}^{1} {\rm d}\alpha{\rm d}\beta
(\alpha\beta)^{\frac{\eta s-\lambda}{\eta}-1}
(\bar{\alpha}\bar{\beta})^{\frac{s+\lambda}{\eta}-1}
\cdot
\Psi_l\bigl[\alpha x_1+\bar{\alpha}x_{2};
\beta x_2 + \bar{\beta} x_{1}\bigr],
$$
so that we obtain:
$$
Q_l(\lambda)= (-1)^l\Gamma^2(\lambda;s)
\int_{0}^{1} {\rm d}\alpha{\rm d}\beta
(\alpha\beta)^{\frac{\eta s-\lambda}{\eta}-1}
(\bar{\alpha}\bar{\beta})^{\frac{s+\lambda}{\eta}-1}
(1-\alpha-\beta)^l.
$$
The eigenvalue $Q_l(\lambda)$ was obtained in
equivalent form in the paper~\cite{F} and
polynomials(in $\lambda$) $Q_l(\lambda)$ coincide
with the Hanh orthogonal polynomials.

\section{Q-operator for $\lambda = \pm \eta s$ and
local Hamiltonians}
\setcounter{equation}{0}

Let us consider the $Q$-operator in $\alpha$-representation:
$$
\hat{Q} (\lambda) \Psi(x) \equiv
\prod_{k=1}^{n}
\Gamma(\lambda;s)\int_{0}^{1} \d\alpha_k
\alpha_k^{\frac{\eta s-\lambda}{\eta}-1}
\bar{\alpha}_k^{\frac{\eta s+\lambda}{\eta}-1}
\Psi\bigl[...\alpha_k x_k+\bar{\alpha}_k x_{k-1}...\bigr]
$$
for the special value of spectral parameter
$\lambda = \eta s + \eta \epsilon$ and calculate the first
two terms of the $\epsilon$-expansion.

We start from the $\alpha_k$-integral:
$$
\frac{\Gamma(2s)}{\Gamma(2s+\epsilon)\Gamma(-\epsilon)}\cdot
\int_{0}^{1} \d\alpha
\alpha^{-\epsilon-1}\bar{\alpha}^{2s+\epsilon-1}
\Psi\bigl[...\alpha x_k+\bar{\alpha}x_{k-1}...\bigr].
$$
The prefactor in this expression is proportional to $\epsilon$ and there
exists the singular $\epsilon$-pole term in the
$\alpha$-integral because of singularity in the point $\alpha=0$.
For the calculation of the $\epsilon$-pole term one can
put $\alpha=0$ in the argument of the $\Psi$-function:
$$
\int_{0}^{1} \d\alpha
\alpha^{-\epsilon-1}\bar{\alpha}^{2s+\epsilon-1}
\Psi\bigl[\alpha x_k+\bar{\alpha} x_{k-1}\bigr] \rightarrow
\frac{\Gamma(-\epsilon)\Gamma(2s+\epsilon)}{\Gamma(2s)}\cdot
\Psi\bigl[x_{k-1}\bigr].
$$
In the main order of $\epsilon$-expansion we need the
singular part of the integral only and have:
\begin{equation}
\label{sh}
\hat{Q} (\eta s) \Psi(x_1,x_2...x_k...x_n) =
\Psi(x_n,x_1...x_{k-1}...x_{n-1}).
\end{equation}
Therefore the $Q$-operator for $\lambda = \eta s$
coincides with the "shift" operator $P$:
$$
P\Psi(x_1,x_2...x_k...x_n) = \Psi(x_n,x_1...x_{k-1}...x_{n-1})
\ ;\ \hat{Q} (\eta s) = P.
$$
In the next order of the $\epsilon$-expansion we have to extract
the $\epsilon$-pole contributions from the $n-1$ $\alpha$-integrals
and the next term of the $\epsilon$-expansion from the one
remaining integral.
This remaining $\alpha_k$-integral has the form:
$$
\frac{\Gamma(2s)}{\Gamma(2s+\epsilon)\Gamma(-\epsilon)}\cdot
\int_{0}^{1} \d\alpha_k
\alpha_k^{-\epsilon-1}\bar{\alpha}_k^{2s+\epsilon-1}
\Psi(...x_{k-2},\alpha_k x_k+\bar{\alpha}_k x_{k-1},x_k...).
$$
Note that $\epsilon$-pole contributions effectively shift all
arguments of the $\Psi(x_1...x_n)$-function except for the $k$-th one.
For the calculation of $\alpha_k$-integral it is useful
to add and subtract the pole term:
$$
\frac{\Gamma(2s)}{\Gamma(2s+\epsilon)\Gamma(-\epsilon)}\cdot
\int_{0}^{1} {\rm d}\alpha
\alpha^{-\epsilon-1}\bar{\alpha}^{2s+\epsilon-1}
\biggl[
\Psi(\alpha x_k+\bar{\alpha}x_{k-1})\pm \Psi(x_{k-1})\biggr].
$$
The integral with the difference is regular so we can put
$\epsilon = 0$ in integrand and extract the needed contribution:
$$
-\epsilon\int_{0}^{1} {\rm d}\alpha
\frac{\bar{\alpha}^{2s-1}}{\alpha}
\biggl[
\Psi(\alpha x_k+\bar{\alpha}x_{k-1})- \Psi(x_{k-1})\biggr]+
\Psi(x_{k-1}).
$$
Finally we obtain the first two terms in the
$\epsilon$-expansion of the Q-operator:
$$
\hat{Q} (\eta s + \eta\epsilon) = P +
\epsilon \sum_{k=1}^{n} H^{-}_{k-1,k} + O(\epsilon^2),
$$
where the operator $H^{-}_{k-1,k}$ is defined as follows:
$$
H^{-}_{k-1,k}
\Psi(x_1...x_{k}...x_n)=
-\int_{0}^{1} {\rm d}\alpha
\frac{\bar{\alpha}^{2s-1}}{\alpha}
\biggl[
\Psi(...x_{k-2},\alpha x_k+\bar{\alpha}x_{k-1},x_k...)-
\Psi(...x_{k-2},x_{k-1},x_k...)\biggr].
$$
Note this "two-particle" operator is not $SL(2)$-invariant.

In a similar way one can calculate the first two terms of
$\epsilon$-expansion for $\lambda = -\eta s$:
$$
\hat{Q} (-\eta s+ \eta\epsilon) = 1 -
\epsilon \sum_{k=1}^{n} H^{+}_{k-1,k} + O(\epsilon^2),
$$
where:
$$
H^{+}_{k-1,k}
\Psi(x_1...x_{k}...x_n)=
-\int_{0}^{1} {\rm d}\alpha
\frac{\bar{\alpha}^{2s-1}}{\alpha}
\biggl[
\Psi(...x_{k-1},\bar{\alpha}x_k + \alpha x_{k-1},x_{k+1}...)-
\Psi(...x_{k-1},x_{k},x_{k+1}...)\biggr].
$$
Let us consider the $\epsilon$-expansion of the
following combination of the Q-operators:
$$
\hat{Q}^{-1}(\eta s) \hat{Q}(\eta s+\eta \epsilon)-
\hat{Q}^{-1}(-\eta s)\hat{Q}(-\eta s+\eta \epsilon) =
\epsilon \sum_{k=1}^{n} H_{k-1,k} + O(\epsilon^2).
$$
Using the expressions for the operators $H^{-}_{k-1,k}$ and
$H^{+}_{k-1,k}$ it is easy to check that operator $H_{k-1,k}$
acts on the variables $z_{k-1},z_k$ only and
coincides to the integral operator considered in~(\ref{Ham}).
Finally we have found the following operator relations:
\begin{itemize}
\item
$\hat{Q} (-\eta s) = 1$
\item $ \hat{Q} (\eta s) = P $
\item
$\hat{Q}^{-1}(\eta s) \hat{Q} (\eta s+ \eta \epsilon) -
\hat{Q}^{-1}(-\eta s)\hat{Q} (-\eta s + \eta \epsilon) =
\epsilon H + O(\epsilon^2)
\ ;\  H \equiv \sum_{k=1}^{n} {\rm H}_{k-1,k}
$
\end{itemize}
Let us compare these relations to the one obtained by
the ABA method~(\ref{HQ},\ref{Pl}).
The first relation fixes normalization of the Q-operator
and the normalization of the eigenvalues of Q-operator:
\be
\label{ql}
Q(\lambda) = \prod_{j=1}^{l} \frac{\lambda-\lambda_j}{-\eta s-\lambda_j}
\ee
The second relation allows to express the eigenvalues of
the "shift" $P$-operator in terms of the function $Q(\lambda)$:
$$
P_l = Q(\eta s) = \prod_{j=1}^{l} \frac{\lambda_j-\eta s}{\lambda_j+\eta s}
$$
in agreement with~(\ref{Pl}).
The thrird relation is the operator version of the equality~(\ref{HQ}).

\section{Asymptotic expansion of Q-operator for $\lambda\to\infty$}
\setcounter{equation}{0}

L.N.Lipatov~\cite{L} has found some beautiful symmetry of the XXX model:
duality transformation.
In this section we show that operator ${\cal S}$ of duality arises
naturally as leading term in asymptotic of Q-operator
for large $\lambda$.

To start with let us define some transformation wich is analogous
to the Fourier tranformation from the coordinate representation to the
momentum representation.

\subsection{Momentum representation}
\setcounter{equation}{0}

Let us define the transformation $T$ from the function $\bar{\Psi}(x)$
in "momentum" representation to the used so far function $\Psi(x)$ in
"coordinate" representation:
$$
\Psi(x)=T\biggl[\bar{\Psi}(x)\biggr]
\ ;\ \Psi(x_1...x_n)\equiv \bar{\Psi}(\partial_{a_1}...\partial_{a_n})
\left. \prod_{k=1}^{n}\frac{1}{[1-a_k x_k]^{2s}}\right|_{a=0}.
$$
This transformation maps polynomials to polynomials and
can be represented as composition of Laplace
transformation and inversion:
$$
T\biggl[\bar{\Psi}(x)\biggr]= \frac{1}{\Gamma(2s)}
R \int_{0}^{\infty}\d t e^{-t x}t^{2s-1}\bar{\Psi}(t).
$$
Using the well known properties of Laplace transformation
and eqs.~(\ref{conj}) it is easy to derive the expression
for the $SL(2)$-generators in "momentum" representation:
$$
T\biggl[x\bar{\Psi}(x)\biggr] =
[x^2\partial + 2s x]\Psi(x)\ ;\
T\biggl[(x\partial^2+2s\partial)\bar{\Psi}(x)\biggr] =
\partial \Psi(x)
$$
$$
T\biggl[(x\partial + s)\bar{\Psi}(x)\biggr]
= (x\partial + s) \Psi(x).
$$
To obtain the rules for the transformation of
commuting operators $Q_k$~(\ref{Qk}) from one representation
to the another we start from the very beginning and consider
transformation of the $L$-operator.
The $L$-operator in coordinate representation
is the T-transformation from the $L^{\prime}$-operator
in "momentum" representation:
$$
L = \left (\begin{array}{cc}
\lambda+\eta[x\partial +s] & -\eta \partial \\
 \eta [x^2\partial+2s\partial] &
\lambda-\eta[x\partial +s] \end{array} \right )
\ ;\  L^{\prime} =\left (\begin{array}{cc}
\lambda+\eta[x\partial+s] & -\eta [x\partial^2+2s\partial] \\
 \eta x & \lambda-\eta[x\partial +s] \end{array} \right )
= \sigma_2 \bar{L} \sigma_2,
$$
where
$$
\bar{L} \equiv \left (\begin{array}{cc}
\lambda-\eta[x\partial +s] & -\eta x \\
 \eta [x\partial^2+2s\partial] &
\lambda+\eta[x\partial +s] \end{array} \right ).
$$
The $\sigma_2$-matrices are cancelled for the transfer matrix
and we can work directly with $\bar{L}$.

Finally we have the formal rules for the transformation from
one representation to the another one:
$$
\eta \rightarrow -\eta \ ;\ x \rightarrow \partial ,
$$
and operators $x$ and $\partial$ have to be "normal ordered":
all $\partial$ stay on the right from the $x$.

\subsection{Asymptotic expansion for large $\lambda$}
\setcounter{equation}{0}

Let us calculate the asymptotic of the $Q$-operator:
$$
\hat{Q} (\lambda) \Psi(x) =
\prod_{k=1}^{n}
\Gamma(\lambda;s)\int_{0}^{1} {\rm d}\alpha_k
\alpha_k^{\frac{\eta s-\lambda}{\eta}-1}
\bar{\alpha}_k^{\frac{\eta s+\lambda}{\eta}-1}
\Psi\bigl[...\alpha_k x_k+\bar{\alpha}_k x_{k-1}...\bigr],
$$
for large values of spectral parameter $\lambda$.
Without lose of generality we can restrict the full
space of polynomials to the subspace of homogeneous~(degree $l$)
polynomials $\Psi(x)$~(\ref{l}).

There exists the expression for Q-operator wich is
more useful for the calculation of the asymptotic:
\be
\label{QPhi}
\hat{Q} (\lambda) \Psi(x) =
\bar{\Psi}(\partial_{a})\prod_{k=1}^{n}
\left. \frac{1}{[1-a_k x_k]^{s-\lambda\eta^{-1}}
[1-a_k x_{k-1}]^{s+\lambda\eta^{-1}}}\right|_{a=0}.
\ee
This formula can be obtained as follows:
$$
\int_{0}^{1} \d\alpha
\alpha^{\frac{\eta s-\lambda}{\eta}-1}
\bar{\alpha}^{\frac{\eta s+\lambda}{\eta}-1}
\Psi\bigl[\alpha x_k+\bar{\alpha}x_{k-1}\bigr]=
\bar{\Psi}[\partial_{a_k}]
\int_{0}^{1} \d\alpha
\frac{\alpha^{\frac{\eta s-\lambda}{\eta}-1}
\bar{\alpha}^{\frac{\eta s+\lambda}{\eta}-1}}
{[1-a_k(\alpha x_k+\bar{\alpha}x_{k-1})]^{2s}}=
$$
$$
=\Gamma^{-1}(\lambda;s)\bar{\Psi}[\partial_{a_k}]
\frac{1}{[1-a_k x_k]^{s-\lambda\eta^{-1}}
[1-a_k x_{k-1}]^{s+\lambda\eta^{-1}}},
$$
where the Feynman formula~(\ref{F}) is used in "opposite" direction.

For the calculation of the asymptotic it is useful to
rescale variables $a_i$ and use the standard expansion for logarithm:
$$
a_i \rightarrow \frac{\eta a_i}{\lambda}
\ ;\ \biggl(1-\frac{a x}{\lambda}\biggr)^{\lambda-s}=
\exp\biggl\{ -ax+\frac{2s a x- a^2 x^2}{2\lambda}+ ...\biggr\}.
$$
Let us consider the contribution with $a_k$:
$$
\bar{\Psi}(\partial_{a_k})
\left.
\frac{1}{[1-a_k x_k]^{s-\lambda\eta^{-1}}
[1-a_k x_{k-1}]^{s+\lambda\eta^{-1}}}\right|_{a_k=0} =
$$
$$
=\bar{\Psi}(\lambda\eta^{-1}\partial_a)
\left.
\exp\biggl\{a(x_{k-1}-x_k)+\frac{\eta(x_{k-1}+x_k)}{2\lambda}
\bigl(2 s a + (x_{k-1}-x_k)a^2 \bigr)+...\biggr\}
\right|_{a=0}=
$$
$$
=\bar{\Psi}(\lambda\eta^{-1}z) +
\frac{\eta(x_{k-1}+x_k)}{2\lambda}
\left.
\bigl(x \partial^2+ 2s\partial\bigr)
\bar{\Psi}(\lambda\eta^{-1}x)\right|_{x=x_{k-1}-x_k} + O(\lambda^{-2})
$$
Polynomial $\Psi(x)$ is homogeneous so that:
$$
\bar{\Psi}(\lambda\eta^{-1}x)
= (\lambda\eta^{-1})^l \bar{\Psi}(x)
$$
and finally we obtain the first two terms of
asymptotic expansion of the Q-operator
for large $\lambda$:
\be
\label{QS}
\hat{Q} (\lambda) = \sum_{k=0}^{l} \hat{Q}_k \cdot(\lambda\eta^{-1})^{l-k}
\ ;\ \hat{Q}_0 \Psi(x) = \bar{\Psi}(x_{n}-x_{1},x_{1}-x_2,...,x_{n-1}-x_{n})
\ee
$$
\hat{Q}_1 \Psi(x) = \frac{1}{2}\sum_{k=1}^{n}
\left.
(x_k+x_{k-1})\bigl(z_k \partial_{z_k}^2+ 2s\partial_{z_k}\bigr)
\bar{\Psi}(z)\right|_{z_k=x_{k-1}-x_k}.
$$
It seems that all operators $\hat{Q}_k$ are local differential
operators in momentum representation.

The operator ${\cal S}$ of duality transformation is defined
in the following way:
$$
{\cal S} \Psi(x_1...x_n) \equiv
\bar{\Psi}(x_n -x_1,x_1-x_2,...,x_{n-1}-x_n).
$$
This operator coincides with the leading term $\hat{Q}_0$
of asymptotic expansion and therefore operator ${\cal S}$ commutes with all
integrals of motion $Q_k$~(\ref{Qk}).
Its common eigenfunction have to be eigenfunction of the ${\cal S}$-operator:
$$
{\cal S} \Psi(x_1...x_n) = {\cal S}_l\cdot \Psi(x_1...x_n)
\Leftrightarrow
\bar{\Psi}(x_n -x_1,x_1-x_2,...,x_{n-1}-x_n) =
{\cal S}_l\cdot \Psi(x_1...x_n)
$$
Note that subspace of the common eigenvectors of $Q_k$ with some
eigenvalues $q_k$ is the $SL(2)$ - module
generated by highest weight vector $\Psi$.
The highest weight vector $\Psi$ is defined by
the equation $S^{-}\Psi = 0$ and can be constructed by
the ABA method~(\ref{S-}).
>From the expression for the eigenvalue of the
Q-operator~(\ref{ql}) one can derives the expression
for the ${\cal S}_l$:
$$
Q(\lambda) =
\prod_{j=1}^{l} \frac{\lambda-\lambda_j}{-\eta s-\lambda_j}
\rightarrow
\frac{(-\lambda)^l}{\prod_{j=1}^{l}(\eta s+\lambda_j)}
\ ;\ \hat{Q}(\lambda)\rightarrow (\lambda\eta^{-1})^l \cdot {\cal S}.
$$
Therefore the eigenvalue of the duality operator for the
highest weight vector of $SL(2)$-module has the form:
$$
{\cal S}_l = \frac{(-\eta)^l}{\prod_{j=1}^{l}(\eta s+\lambda_j)}.
$$
It is easy to see that all other vectors from $SL(2)$-module
form the "zero" subspace:
$$
\Phi(x_1...x_n) = S^{+}\Psi(x_1...x_n)\rightarrow
{\cal S}\Phi(x_1...x_n) =
{\cal S} (x_1+...+x_n)\bar{\Psi}(x_1...x_n) = 0.
$$

\section{Conclusions.}

We have constructed the Baxter's Q-operator for
the homogeneous XXX spin chain and have checked the
consistence of obtained results with the corresponding
formulae obtained in the framework of the ABA-method.
We have found the connection between Q-operator and
operator of duality symmetry.

The considered construction can be applied to the
inhomogeneous XXX-model but we are not able to obtain
the useful and compact representation for the
Q-operator in this case.

There exists some more universal approach to the construction of
the quantum Q-operator.
As V.B.Kuznetsov and E.K.Sklyanin informed me,
they have recently obtained similar results~\cite{KS1} using the
approach of~\cite{KS} based on correspondence between the quantum
Q-operator and classical B\"acklund transformation.

\subsection*{Acknowledgments}

I am very grateful to I.V.Komarov, V.B.Kuznetsov, L.N.Lipatov and
E.K.Sklyanin for their interest in the work and useful discussions.
I would like to thank especially A.N.Manashov for
critical remarks and stimulating disscussions.
This work was supported by RFFR Grant 97-01-01152 of the RFFR.

\setcounter{equation}{0}
\renewcommand{\theequation}{B.\arabic{equation}}
\section*{Appendix}

In this Appendix we prove the identity:
$$
\langle (1 - x_{k-1} y)^{-2s+a}(1- x_k y)^{-a}|
(1- y z_{k})^{-b}(1-y z_{k+1})^{-2s+b}\rangle
\cdot (1- z_k x_{k-1})^{a-b}=
$$
\be
\label{local}
=
\langle
(1- x_{k-1} y)^{-2s+b}(1- x_k y)^{-b}|
(1 - y z_{k})^{-a}(1-y z_{k+1})^{-2s+a}\rangle
\cdot (1-z_{k+1}x_{k})^{a-b}
\ee
First of all note that the equality~(\ref{id}) can be
rewritten in integral form:
$$
\langle (1-x_{k-1}z)^{-a}(1-x_{k-1}z)^{-b}|\Psi(z)\rangle
=\frac{\Gamma(2s)}{\Gamma(a)\Gamma(b)}
\frac{1}{(x_k-x_{k-1})^{2s-1}}\cdot
$$
$$
\cdot\int_{x_{k-1}}^{x_{k}} {\rm d}t
(t-x_{k-1})^{b-1}(x_k-t)^{a-1}\Psi(t)
$$
and therefore the identity~(\ref{local}) is equivalent
to the following identity for integrals:
$$
\frac{\Gamma(2s)}{\Gamma(2s-a)\Gamma(a)}
\frac{(1- x_{k-1}z_k)^{a-b}}{(x_k-x_{k-1})^{2s-1}}\cdot
\int_{x_{k-1}}^{x_{k}} {\rm d}t
\frac{(t-x_{k-1})^{a-1}(x_k-t)^{2s-a-1}}
{(1- t z_{k})^{b}(1-t z_{k+1})^{2s-b}}=
$$
\begin{equation}
\label{integr}
=
\frac{\Gamma(2s)}{\Gamma(2s-b)\Gamma(b)}
\frac{(1- x_{k}z_{k+1})^{a-b}}{(x_k-x_{k-1})^{2s-1}}\cdot
\int_{x_{k-1}}^{x_{k}} {\rm d}\tau
\frac{(\tau-x_{k-1})^{b-1}(x_k-\tau)^{2s-b-1}}
{(1- \tau z_{k})^{a}(1-\tau z_{k+1})^{2s-a}}.
\end{equation}
Let us start from the t-integral.
There exists the bilinear transformation with
the properties:
$$
z= S x = \frac{A x-C}{C x+D}
\ ;\  S x_k = z_k\ ,\ S x_{k-1} = z_{k+1}
$$
This transformation can be obtained as follows:
$$
\frac{z-z_{k}}{z-z_{k+1}}=R\frac{x-x_{k}}{x-x_{k-1}}
\Rightarrow
z=\frac{x(x_k-x_{k-1}R)+ z_k x_{k-1}R-x_k z_{k+1}}
{x(1-R)+z_kR-z_{k+1}} = \frac{A x-C}{C x+D}
$$
and therefore:
$$
A=(x_k-x_{k-1}R)\ ;\ D=z_kR-z_{k+1}\ ;\ C=1-R
\ ;\ R=\frac{1-x_k z_{k+1}}{1-z_k x_{k-1}}
$$
It is worth to emphasize the additional properties:
$$
S z_k = x_k\ ,\ S z_{k+1} = x_{k-1}
\ ;\  \frac{C z_{k+1}+ D}{C z_k +D} = R
$$
Let us make the same bilinear transformation in $t$-integral:
$$
t = S\tau = \frac{A \tau-C}{C \tau+D}
\ ,\ 1-t z_k = \frac{Cz_k+D}{C\tau+D}(1-\tau x_k)
\ ;\ x_k= S z_k \ ,\ x_{k-1} = S z_{k+1}.
$$
Then we obtain:
$$
\frac{1}{(x_k-x_{k-1})^{2s-1}}\cdot
\int_{x_{k-1}}^{x_{k}} {\rm d}t
\frac{(t-x_{k-1})^{a-1}(x_k-t)^{2s-a-1}}
{(1- t z_{k})^{b}(1-t z_{k+1})^{2s-b}}=
$$
$$
=\frac{(Cz_{k+1}+D)^{a-b}}{(Cz_{k}+D)^{a-b}}
\frac{1}{(z_k-z_{k+1})^{2s-1}}\cdot
\int_{z_{k+1}}^{z_{k}} {\rm d}\tau
\frac{(\tau-z_{k+1})^{a-1}(z_k-\tau)^{2s-a-1}}
{(1- \tau x_{k})^{b}(1-\tau x_{k-1})^{2s-b}}
$$
On the next step we transform the $\tau$-integral using
the $\alpha$-representation and the Feynman formula:
$$
\frac{1}{(z_k-z_{k+1})^{2s-1}}\cdot
\int_{z_{k+1}}^{z_{k}} {\rm d}\tau
\frac{(\tau-z_{k+1})^{a-1}(z_k-\tau)^{2s-a-1}}
{(1- \tau x_{k})^{b}(1-\tau x_{k-1})^{2s-b}}=
$$
$$
=\int_{0}^{1} {\rm d}\alpha
\frac{\alpha^{a-1}\bar\alpha^{2s-a-1}}
{[1- (\alpha z_k+\bar\alpha z_{k+1}) x_{k}]^{b}
[1-(\alpha z_k+\bar\alpha z_{k+1}) x_{k-1}]^{2s-b}}=
$$
$$
=\frac{\Gamma(2s)}{\Gamma(b)\Gamma(2s-b)}
\int_{0}^{1} {\rm d}\beta \beta^{b-1}\bar\beta^{2s-b-1}
\int_{0}^{1} {\rm d}\alpha
\frac{\alpha^{a-1}\bar\alpha^{2s-a-1}}
{[1- (\alpha z_k+\bar\alpha z_{k+1})
(\beta x_k+\bar\beta x_{k-1}) ]^{2s}}=
$$
$$
=\frac{\Gamma(a)\Gamma(2s-a)}{\Gamma(b)\Gamma(2s-b)}
\int_{0}^{1} {\rm d}\beta \frac{\beta^{b-1}\bar\beta^{2s-b-1}}
{[1- (\beta x_k+\bar\beta x_{k-1}) z_{k}]^{a}
[1-(\beta x_k+\bar\beta x_{k-1}) z_{k+1}]^{2s-a}}=
$$
$$
=\frac{\Gamma(a)\Gamma(2s-a)}{\Gamma(b)\Gamma(2s-b)}
\frac{1}{(x_k-x_{k-1})^{2s-1}}\cdot
\int_{x_{k-1}}^{x_{k}} {\rm d}\tau
\frac{(\tau-x_{k-1})^{b-1}(x_k-\tau)^{2s-b-1}}
{(1- \tau z_{k})^{a}(1-\tau z_{k+1})^{2s-a}}.
$$
Collect all together we obtain the equality~(\ref{integr}).

\end{document}